# A Redundancy-Guided Approach for the Hazard Analysis of Digital Instrumentation and Control Systems in Advanced Nuclear Power Plants

**Tate Shorthill[*], Han Bao[†], Hongbin Zhang[†], Heng Ban[*]**

[*]University of Pittsburgh, 3700 O'Hara Street, Pittsburgh, PA 15261
[†]Idaho National Laboratory, 2525 Fremont Ave, Idaho Falls, ID 83402

## ABSTRACT

Digital instrumentation and control (I&C) upgrades are a vital research area for nuclear industry. Despite their performance benefits, deployment of digital I&C in nuclear power plants (NPPs) has been limited. Digital I&C systems exhibit complex failure modes including common cause failures (CCFs) which can be difficult to identify. This paper describes the development of a redundancy-guided application of the Systems-Theoretic Process Analysis (STPA) and Fault Tree Analysis (FTA) for the hazard analysis of digital I&C in advanced NPPs. The resulting Redundancy-guided System-theoretic Hazard Analysis (RESHA) is applied for the case study of a representative state-of-the-art digital reactor trip system. The analysis qualitatively and systematically identifies the most critical CCFs and other hazards of digital I&C systems. Ultimately, RESHA can help researchers make informed decisions for how, and to what degree, defensive measures such as redundancy, diversity, and defense-in-depth can be used to mitigate or eliminate the potential hazards of digital I&C systems.

## ABBREVIATIONS

| | |
|---|---|
| AOO | anticipated operational occurrence |
| BP | bistable processor |
| CCF | common cause failure |
| CS | control structure |
| D3 | diversity and defense-in-depth |
| DOE | U.S. Department of Energy |
| DPS | diverse protection system |
| EPRI | Electric Power and Research Institute |
| ESFAS | engineered safety feature actuation system |
| FMEA | Failure Mode and Effects Analysis |
| FT | fault tree |
| FTA | Fault Tree Analysis |
| HAZCADS | Hazard and Consequence Analysis for Digital Systems |
| I&C | instrumentation and control |
| INL | Idaho National Laboratory |
| LC | logic cabinet |
| LCL | local coincidence logic |
| LP | logic processor |
| MCR | main control room |
| NPP | nuclear power plant |
| NRC | U.S. Nuclear Regulatory Commission |
| RESHA | Redundancy-guided Systems-theoretic Hazard Analysis |
| RPS | reactor protection system |

| RSR | remote shutdown room |
| RTB | reactor trip breaker |
| RTS | reactor trip system |
| SOI | system of interest |
| SP | selective processor |
| SPOF | single point of failure |
| ST | shunt |
| STPA | Systems-Theoretic Process Analysis |
| UCA | unsafe control action |
| UV | undervoltage |

# 1. INTRODUCTION

Digital upgrades and plant modernization efforts offer the foremost path to performance and cost improvements of nuclear power plants (NPPs) [1]. Despite decades of experience with analog systems, the technical challenges associated with their continued use (e.g., signal drift, high maintenance costs, obsolescence, and lack of industrial suppliers) have caused the nuclear industry to move toward digital instrumentation and control (I&C) in favor of integrated circuitry and the modern microcontroller [2] [3]. Compared with analog, digital I&C systems offer significant advantages in the areas of monitoring, processing, testing, and maintenance [2] [4]. Notwithstanding the immediate attraction, the nuclear industry has been slow to adopt safety rated digital I&C because each new design must be shown to maintain or improve the status quo by means of a risk assessment [3]. Though many of the concepts for the risk assessment of analog systems carry over, digital I&C systems present unique challenges. In 1997, the National Research Council detailed several technical challenges for the implementation of digital I&C systems. Those relating specifically to the present work are: (1) the system aspects of digital systems; (2) the potential for software-based common cause failures (CCFs); and (3) the need for a risk assessment method tailored to digital I&C systems [3].

The system aspects of digital I&C involve issues that extend beyond individual components and even beyond the function of the system itself. The challenge of the system aspects is discussed in NUREG/CR-6901. Digital systems exhibit two types of interactions—Type 1: the interactions of a digital I&C system (and/or its components) with a controlled process (e.g., NPP); and Type 2: the interactions of a digital I&C system (and/or its components) with itself and/or other digital systems and components [5]. Kirschenbaum et al. provide a useful summary of these concerns in their own work in the investigation of digital systems [6]. Additionally, NUREG/CR-6734 Vol.1 & Vol.2 provide a list of guidelines for software systems requirements and an expansive list of real failure examples relating to software [7] [8]. A chief concern for software failures are those effecting common systems or software (i.e., CCFs).

Common or redundant components are often utilized as a backup to ensure system reliability. However, the improper application of redundant features can leave a system vulnerable to CCFs, which arise from the malfunction of two or more components, or functions, due to a single failure source [1] [9]. In order to make redundancy designs effective, diversity is employed, providing an alternative technology, method, technique, or means to achieve a desired result [10]. The diverse protection helps to eliminate the common features necessary for a CCF. NUREG/CR-5485 provides general guidance for modeling CCFs in risk assessments [11]. While, NUREG/CR-6303 is written in the context of software systems and provides guidance for the application of diversity to minimize CCFs [12]. These reports provide a basis for dealing with CCFs. There need remains to identify the most significant CCFs in order to focus the application of diversity in design.

While diversity in analog I&C systems is a standard practice, its effectiveness for digital I&C systems remains a topic of concern. The use of digital technologies may "increase the potential for CCF vulnerabilities because of the introduction of undetected systematic faults" [13]. These systematic failures (i.e. failures inherent to design specifications and systems interactions [14]) in the software may result from, "(1) errors in the requirement specification, (2) inadequate provisions to account for design limits (e.g., environmental stress), or (3) technical faults incorporated in the internal system (or architectural) design or implementation" [13]. Knight and Leveson indicate that independently developed software may not necessarily fail independently [15]. These sources highlight the importance of properly defined software requirements to help eliminate system failures. To effectively determine software requirements, one must identify system vulnerabilities. As summarized by Arndt and Kuritzky in 2011, the identification and modeling of software failure modes remains the most challenging aspect of risk analysis for digital I&C systems [16]. Once system vulnerabilities have been identified, NUREG/CR-6303 and NUREG/KM-0009 can provide approaches on how to limit them [12] [17]. More recently, NUREG/CR-7007 was written to determine "how much diversity is enough" [18].

The implementation of digital I&C upgrades depends on risk assessment methods which can appropriately handle the unique aspects and CCFs of digital systems. A proper risk assessment relies on both qualitative and quantitative techniques [11]. In 1997, The National Research Council indicated concern within the software community whether failure rate probabilities or quantification is even appropriate for software [3]. Reliability assessments of digital systems have remained a significant focus in the literature; NUREG/CR-6901, NUREG/CR-6942, and NUREG/CR-6985 are just a few examples [5] [19] [20]. The hazard analysis performed in this work provides a technical basis for a reliability analysis by identifying crucial failure modes and qualitatively determining their effects on system vulnerability.

To support the transition from analog to digital I&C systems, Idaho National Laboratory (INL) has initiated a project under the Risk-Informed Systems Analysis (RISA) Pathway of the U.S. Department of Energy's (DOE's) Light Water Reactor Sustainability program to develop a risk assessment strategy for the design and use of digital I&C upgrades [21]. This effort aims to provide a process for the nuclear industry to systematically assess and evaluate the risks of advanced and highly redundant digital systems by answering three questions used by the Nuclear Regulatory Commission (NRC) to define risk: "(1) What can go wrong? (2) How likely is it? (3) What are the consequences?" [22]. These questions are the foundation of the risk assessment strategy being developed by INL. Part 1 corresponds with the identification of hazards. Part 2 involves establishing the frequency or probability of the hazards (i.e., the reliability of the system). Part 3 focuses on evaluating the consequences of the hazards found and quantified from Parts 1 and 2. This paper details the hazard analysis of digital I&C systems with an emphasis on the identification of potential CCFs as Part 1 of the risk assessment strategy [21].

This work provides recommendations for the management of risks and the development and deployment of advanced digital I&C technologies in the nuclear industry by meeting the following objectives: (1) provide a technical basis for the implementation of a reliability analysis by identifying crucial failure modes and qualitatively determining their effects on system vulnerability; (2) help utilities optimize the use of diversity attributes in a cost-effective manner; (3) help engineers efficiently make design and risk mitigation decisions by providing them a means to systematically identify the most critical CCFs and hazards of digital I&C systems.

Section 2 of the paper provides detail regarding technical background and approach to a Redundancy-guided Systems-theoretic Hazard Analysis (RESHA). Next, a detailed description of each step of RESHA is provided in Section 3. RESHA is then demonstrated in a case study hazard analysis of a digital reactor trip system in Section 4. The results are discussed in Section 5, followed by a conclusion of the work in Section 6.

## 2. TECHNICAL BACKGROUND AND APPROACH

Hazards can be understood as a state or condition that can lead to a loss of something of value to stakeholders [23]. Hazard analysis is defined as the process of examining a structure, system, or component in order to identify hazards and triggers of hazards with the goal of eliminating, mitigating or controlling them [24]. The 2013 Electric Power Research Institute (EPRI) report on hazard analysis methods for digital I&C discusses common methods and indicates their respective strengths and weaknesses [25]. In order to successfully model digital I&C systems, the need exists to model both the hardware and software interactions of the system. Traditional methods such as Failure Modes and Effects Analysis (FMEA) and Fault Tree Analysis (FTA) have been used to extensively to model analog systems. However, interactions between digital systems and controlled processes (i.e. Type 1 interactions) and the interactions between digital systems and their own components or other systems (i.e. Type 2 interactions) can result in failure modes or hazards that are difficult to discover using traditional methods [16]. Lessons learned from the NRC's investigation of multiple analysis methods indicate that there "may not be one preferable method for modeling all digital systems" [16]. However, combinations of methods may prove beneficial. A recent advancement in hazard analysis, developed jointly by EPRI and Sandia National Laboratory, combines FTA and systems-theoretic process analysis (STPA) as a portion of their methodology of Hazard and Consequence Analysis for Digital Systems (HAZCADS) [26]. The current work incorporates this concept of combining FTA and STPA as part of the approach for RESHA.

FTA is a conventional, top-down, risk assessment tool that is used to identify the faults that contribute to the failure of a selected top event, which may depend on combinations of multiple smaller contributors known as basic events. A key aspect of FTA is the determination of the "cut set," which is collection of events that, when combined, will result in failure of the selected top event. There are often many variations of cut sets comprising of single or multiple events [27]. For a hazard analysis, the determination of the cut set is critical and can be done without probability, or failure rate, data [28]. FTA is the workhorse of the nuclear industry and has been used extensively to model control systems.

STPA is an analysis method based on system theory that is used to capture the unsafe interactions between system components, in addition to component failures [23]. Its top-down analysis focuses on the identifying constraints on behavior and the interaction of components [29]. The main parts of STPA are focused on the construction of a control structure and the analysis of unsafe component interactions. The control structure highlights the controller interactions, while the analysis identifies necessary behavior constraints to reduce or eliminate hazards. Together FTA and STPA will be used to provide a clearer picture of the hazards in digital I&C systems as shown in Figure 1.

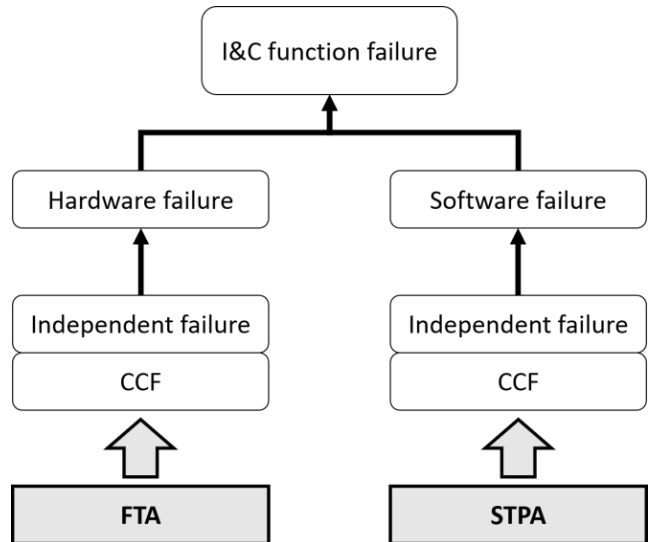

Figure 1. FTA and STPA combined for a full picture of I&C failures adapted from [21].

Though STPA may be applied at any stage of system design and review, it is ideally suited for early application in the design process before safety features have been incorporated into the design [23]. Then, as more details are incorporated, the STPA method is applied iteratively to further improve the design. However, even when fine detail about a system is known, the analysis may remain at a high level, relying on causal factor investigations to provide the detail of subcomponent failures and interactions. In other words, the details such as redundant subsystems or components are often ignored in all but the final part of STPA. Consequently, there isn't a clear representation of how to apply STPA or create a control structure for a system containing multiple layers of redundancy. In a 2014 Massachusetts Institute of Technology technical report, STPA is applied to an aircraft automatic braking system and redundant features are only mentioned in the context of causal factors [29]. In 2018, Rejzek and Hilbes replicated this idea for an application of STPA to nuclear related digital I&C systems. Their application stresses that the control structure only incorporate the functional aspects of the system, again leaving the details of redundancy for the context portion of STPA [30]. While this aspect of STPA does not appear to lessen its effectiveness, waiting to include the finer details of a system may hinder the identification of CCFs. As a possible consequence of this complication, STPA does not appear to be used specifically to identify and address CCFs in the context of digital I&C.

In order to identify CCFs, including their relationship to characteristics not clearly featured in the standard STPA analyses, it is essential that STPA have a clear application process for highly redundant systems. For this particular application of STPA, all layers of a redundancy design should be incorporated clearly and early in the design analysis. It is proposed that reframing STPA in a redundancy-guided way, in combination with FTA, will provide a means to effectively identify the unique hazards, failure modes and CCFs associated with highly redundant safety-based digital I&C systems [31]. By following this approach, the triggers of the most significant hazards to advanced digital I&C systems can be identified and either mitigated or eliminated.

## 3. REDUNDANCY-GUIDED SYSTEMS-THEORETIC HAZARD ANALYSIS

RESHA includes seven steps that are illustrated in Figure 2. RESHA is not organized in a way to direct the initial concept and genesis of a control structure. Rather, the steps are arranged for the analysis of existing designs. RESHA is ideally suited for the analysis of designs that already have some redundancy features incorporated.

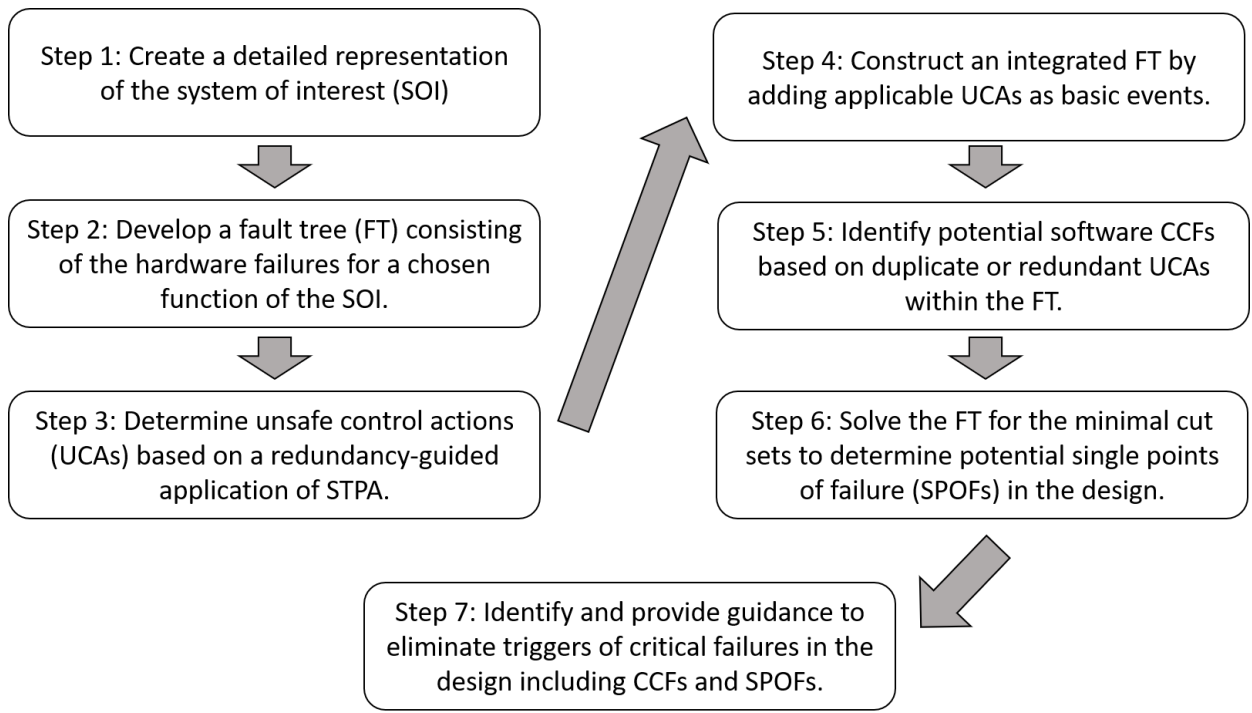

Figure 2: Steps to the redundancy-guided system-theoretic hazard analysis.

**STEP-1: Create a detailed representation of the system of interest (SOI).**

The purpose of STEP-1 is to establish a system sketch to serve as a blueprint for the hazard analysis. This is done by gathering system design information, including wiring, piping and instrumentation diagrams, existing logic diagrams, etc. The information is then used to create a system sketch, the main goal of which is to map out the processors, sensors, controllers, components, interactions, and connections of the system. The point of this step is not to necessarily fit everything into one diagram, but to gain a sufficient understanding of the system in order to complete the hazard analysis; the level of detail provided in this step lays out the foundation for the work.

Digital I&C structures can be divided into three hierarchical levels: (1) divisions; (2) units; and (3) modules. Divisions typically fulfill a role of monitoring processes, systems, or functions redundantly with other divisions by relying on their subsystems of units and modules. Units perform specific tasks that are supported by the function processing capabilities of modules [28]. The system diagram should clearly display the flow of information between divisions, units, and modules. Emphasis should be placed to clearly illustrate the redundancy and diversity of the system in order to provide a basis for constructing the FT in STEP-2 and control structure in STEP-3.

Several key points are considered for STEP-1:

- STEP-1 emphasizes the boundary conditions and scope of the analysis. These should be clearly understood as they will be revisited in STEP-2 and STEP-3.
- Though the RESHA has been developed to analyze digital systems, this system sketch should also include the hardware structural arrangement (i.e., the components of the system in addition to details collected for the digital structural arrangement).

- The level of detail included in a hazard analysis can extend beyond module level failures to the components and sensors that provide input to process modules. The level of detail to be included in the hazard analysis depends on the scope of the investigation.

**STEP-2**: **Develop a fault tree (FT) consisting of the hardware failures for a chosen function of the SOI.**

Having constructed a sketch of the digital system, a corresponding FT can be made. The main assumption for this step is that all software failures will be captured using STPA in STEP-3. Therefore, only hardware failures will be included in FT. The FT is created using the following 2-part process[1] adapted from the NASA Handbook [27].

STEP-2A: Define the boundary of the analysis (revisited from STEP-1). This includes selecting a top event and resolution for the analysis. Top events[2] are based on the purpose of the SOI. The failure of the SOI's most significant function is a priority event to be analyzed by the FT—a top event. STEP-3A may also be visited briefly to ensure the proper selection of top events.

STEP-2B: Construct the FT. Starting from the top event, construction proceeds by determining the "necessary and sufficient immediate events" contributing to failure of the top event [27]. This process is repeated down the tree by analyzing each subsequent event step-by-step, ending with an event that can be resolved no further (either by way of fact, or by the discretion of the analyst) [27]. This event is known as a basic event. The failure of each event depends on a logical combination (e.g., "and," "or," and "n/m" functions) of basic events, which should be built into the FT based on the system diagram from STEP-1. The structure of the FT should capture the details of redundancy that will aid in the subsequent steps of the hazard analysis. In most cases, the highest level of redundancy will be associated with protecting the main function of the SOI. For instance, a system may have two or more divisions, independent and redundant in function, to ensure the reliability of that system. Redundancy also extends to the units and modules of digital systems, for which commonality becomes a possible source of CCF. The hardware CCFs should also be included in the FT. Software details will be added in STEP-3 of the RESHA.

**STEP-3: Determine unsafe control actions (UCAs) based on a redundancy-guided application of STPA.**

The main purpose of STEP-2 is to construct a FT that can be readily combined with software failure events which will be found using a redundancy-guided application of STPA here in STEP-3. According to the STPA Handbook there are four main parts to the STPA. The first two focus on the construction of a control structure, and the second two focus on the analysis of unsafe component interactions. The first three, shown in Figure 3, will be used in STEP-3. The fourth part of STPA is included later in STEP-7. A fully detailed description of the process and steps of STPA can be found in the STPA Handbook 2018 [23]. The process given here highlights the key parts directly applicable to RESHA.

---

[1] This process has been simplified to two parts and directed to the analysis of highly redundant digital I&C from the original eight-step FTA given in the NASA FTA Handbook [27].
[2] There can be multiple top events for a single system with the FT for each new top event varying significantly. The number of FT to be analyzed will depend on the project scope.

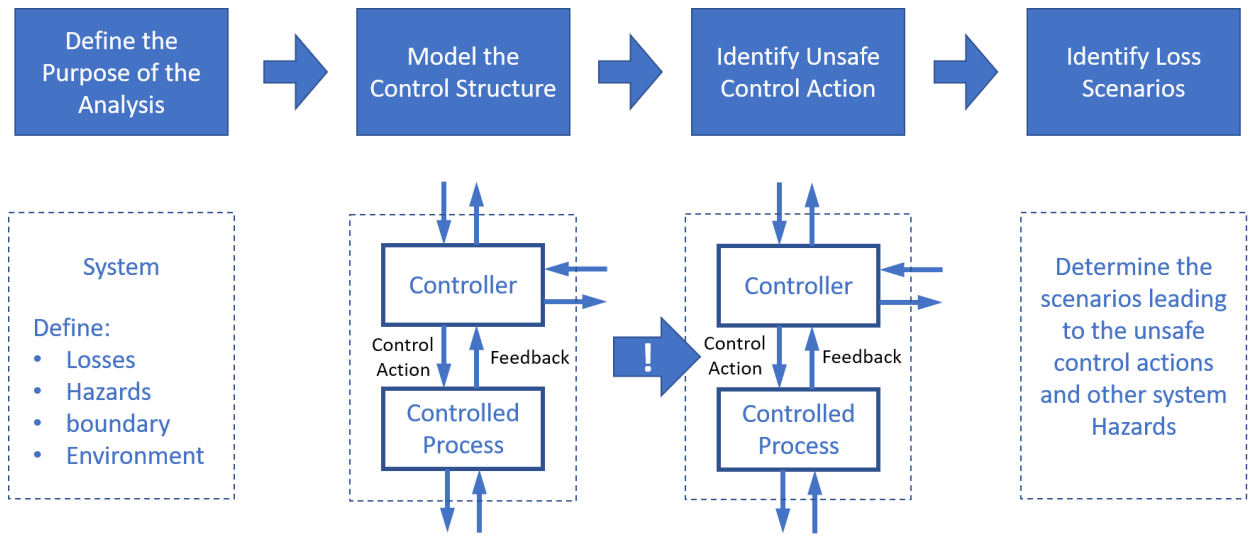

Figure 3. The first three parts of STPA are used in STEP-3 of the RESHA. The details shown have been adapted from the STPA Handbook [23].

STEP-3A: Defining the purpose of the analysis. The SOI should be defined along with its losses, hazards, boundaries, and environments. The system diagram developed in STEP-1 should provide clear indication of the boundary, and environment. A loss is defined as: "something of value to stakeholders. Losses may include a loss of human life or human injury, property damage, environmental pollution, loss of mission, loss of reputation, loss or leak of sensitive information, or any other loss that is unacceptable to the stakeholders" [23]. Losses should be listed to provide direction for identifying hazards. System hazard is a state or set of conditions that, in a particular set of worst-case conditions, will lead to a loss [23]. These hazards pertain to the system as a whole and do not represent a complete hazard analysis.

STEP-3B: Create a model of the control structure. This is done by referring to the system diagram from STEP 1 to identify, group, and organize the system based on its controllers, components and controlled process. The STPA Handbook promotes the idea of "zooming-in" to add system detail [23]. However, as mentioned previously, the typical application of STPA does not build multi-layered redundancy directly into the control structure. By reframing the standard STPA approach, complex and highly redundant digital systems can be broken down into more manageable portions. The redundancy-guided approach allows a straightforward creation of a multi-layer control structure which both captures the details of the SOI and provides an efficient way to identify CCFs.

The first control structure should be based on the highest level of redundancy (e.g., divisional), followed by any sub-layers (e.g., unit and module-based[3] redundancies). Each control structure layer is created with numbered control actions and feedback signals until a final, redundancy-guided, multi-layer control structure is created for the complete SOI, as shown in Figure 4. A table should be used to track each of the identified control actions.

---

[3] Creating a control structure for the internal structure of a module may not be required based on the assumed/desired model resolution and scope.

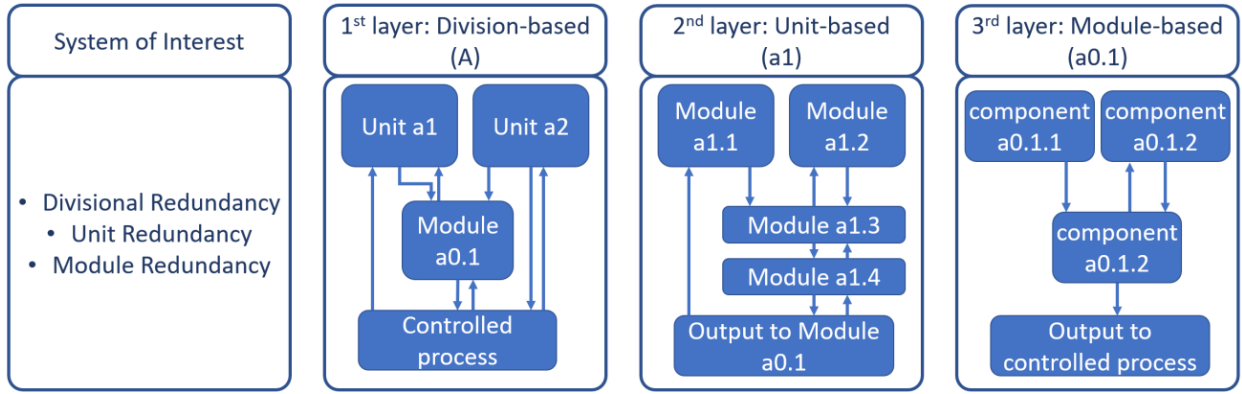

Figure 4:Multi-layered control structure approach. The complete system is modeled using control structures identified and separated according to redundancy. The numbering system (XXnn.nn.nn) corresponds to DIVISIONunit.module.component.

STEP-3C: Identify UCAs. A UCA is control action that, in a particular context and environmental conditions, will lead to a hazard [23]. When the control structure has been completed, the UCAs can be determined from the control action table created in STEP-3B and then compiled in a new table. The UCAs will be used to add software failures to the FT from STEP-2. For each control action identified, there are four categories of potential UCAs that may affect the behavior of the SOI. These categories are:

1. Control action is not provided when it is needed.
2. Control action is provided when it is not needed.
3. Control action is provided when it is needed, but too early, too late, or in a wrong order.
4. Control action lasts too long or stops too soon (only applicable to continuous control actions).

Each UCA should take the following format for consistency [23]:

$$UCA = [source] + [UCA\ type] + [Control\ Action] + [Context] + [Link\ to\ System\ Hazards]$$

**STEP-4 Construct an integrated FT by adding applicable UCAs as basic events.**

This step completes the FT started in STEP-2 by adding applicable UCAs from those found in STEP-3. Applicable UCAs are selected based on their relation to the top event chosen in STEP-2. For example, control actions that are continuous, like those relating to category four above, are only applicable to top events requiring continuous control. Each UCA must be added to the FT as a single basic event. This should be placed in the FT by adding a software failure subtree where appropriate. This will provide a location to add software CCFs in STEP-5. Figure 5 provides an example of this structure.

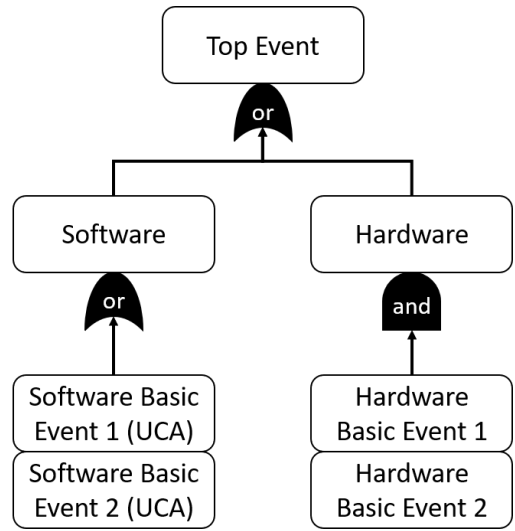

Figure 5. Example FT format for incorporating software and hardware-based failures.

**STEP-5 Identify potential software CCFs based on duplicate or redundant UCAs within the FT.**

The identification of software CCFs depends on the software diversity of the digital system. As one of the guidelines for software, the NRC prescribes the use of diversity and defense-in-depth (D3) as, "the use of different programs designed and implemented by different development groups with different key personnel to accomplish the same safety goals-for example, using two separately designed programs to compute when a reactor should be tripped" [12]. The degree to which D3 is employed can limit the number or potential for CCFs in a control system. Before the identification of software CCFs, the level of system diversity in the SOI should be determined or expressly assumed to guide the classification of software CCFs.

Each software failure can result from for one or more types of UCA; therefore, each component or processor has the potential for multiple software failure events. Additionally, when components share commonality in software or are complete duplications for the sake of redundancy, there is a potential for software CCFs. These CCFs can be grouped according to their hierarchy (i.e., divisions, units, and modules). A module can share a CCF with common modules of its own unit, with common modules of other units, or with common modules across divisions. The table of UCAs and the integrated FT can help identify the CCFs. Based on the research assumptions, the number of CCFs to be included in the final FT can be limited (e.g., to a single division or unit). However, the full set provides a useful tool to direct D3, improve the SOI, and mitigate the hazards of CCFs. Each selected CCF should be placed within the FT where appropriate, such as next to their single failure counterparts.

**STEP-6: Solve the FT for the minimal cut sets to determine potential single points of failure (SPOFs) in the design.**

FTs can be analyzed quantitatively or qualitatively [28]. RESHA relies on the qualitative analysis based purely on the logic of the FT, which identifies crucial failures (i.e., hazards) and provides the technical basis for the reliability (i.e., quantitative) analysis to performed in future work. The qualitative analysis of the FT should be conducted using an FTA software package which will evaluate the FT for minimal cut sets (i.e., a collection of basic events that, when combined, will result in failure of the top event). The number of basic events within each cut set (i.e., cut set order) varies depending on the SOI. For a qualitative analysis, the significance of each cut set is assumed to be dependent only on order; therefore, the fewer basic events in the cut set, the more likely it will occur[4]. For this analysis, a first order cut set is known as a SPOF. Cut sets containing a CCF as their only basic event are also treated as a SPOF for the sake of this analysis.[5] By evaluating the cut set list for SPOFs, designers can determine where and how the SOI is vulnerable and redesign it accordingly.

The next step in the RESHA continues by evaluating the cut sets found in STEP-6. There will be instances where, due to the system design, there are not SPOFs to analyze. In such cases, the analysis should proceed with the next lowest order. The idea is to focus on the cut sets that are most critical based on the assumption that probabilities are irrelevant and higher order cut sets occur less frequently. The FT cut set list can also be modified by evaluating portions of the FT that are most concerning to designers, such as individual divisions or subsystems. By doing so, SPOF for subsystems may be found and addressed.

**STEP-7: Identify and provide guidance to eliminate triggers of critical failures in the design including, CCFs and SPOFs.**

The purpose of the previous six steps was to help efficiently identify the most significant hazards of the SOI. The intention of STEP-7 is to identify and eliminate the triggers or causal factors of the basic events (i.e., UCAs, CCFs, etc.) contained in the cut sets of the FT. The most significant focus here should be to identify the causal factors for the CCFs and SPOFs of the system. The STPA Handbook indicates that causes of UCAs can be grouped into two categories: (1) unsafe controller behaviors; and (2) inadequate feedback or other inputs [23]. The two categories are expanded to include some examples:

- Category 1: Failures related to the controller itself (physical failures); inadequate control algorithms (the decision-making process may be inappropriate); inadequate process model (the controller's view of the controlled process may be incorrect) [23].

- Category 2: Information/Feedback is *not* received by the controller; inadequate feedback/information *is* received by the controller; unsafe control input (a controller receives an UCA input from another controller) [23].

The categories serve as a starting point for analyzing the results from the previous steps. However, the impact of the results of STEP-7 will depend largely on the knowledge and skill of the analysis team for determining the details of the causal factors. As each causal factor is identified, efforts can be made to evaluate or mitigate them by applying D3 or redesigning the system. The initial efforts in identifying causal factors may result in a resource bank of typical causal factors that can be used to expedite future analyses, thus reducing the cost of the RESHA over time.

## 4. CASE STUDY

This section describes the hazard analysis of a four-division digital reactor trip system (RTS), which has a similar structure to state-of-the-art digital systems in existing NPPs [32]. The analysis follows the 7-step process outlined in the preceding section. The following is a list of assumptions built into the analysis:

1. All components of the RTS are assumed to be digital, and are therefore susceptible to software failures. The reactor trip breakers (RTBs) are an exception; they are assumed to be analog, having only hardware failures.

---

[4] The addition of probability values can lead to higher order cut sets occurring with greater probability than those of lower order. These effects will be investigated as part of the reliability analysis in a future work.
[5] According to NRC interim staff guidance on digital systems, CCFs are not considered single failures [33].

2. The functional of an RTB is controlled by two mechanisms, an undervoltage (UV) trip mechanism and a shunt (ST) trip mechanism; the reactor protection system (RPS) activates the UV while the diverse protection system (DPS) activates the ST. The main control room (MCR) and the reserve shutdown room (RSR) are assumed to have an alternate method that causes the RTBs to open.

3. The MCR and the RSR only need to make a single control action to trip the reactor. They do not trip individual divisions.

4. The controller in MCR/RSR is an operator. The UCAs associated with the human controllers are added to the FT in the same manner as the UCAs associated with the software failures of digital controllers.

5. The number of CCFs in the FT is limited to common failures within a single division, and CCFs across all divisions. No CCFs are provided for combinations of divisions. Limiting the combinations of CCFs, in the manner specified, simplified the model without removing the most significant SPOFs caused by CCFs. Assuming failures had to occur across all divisions ensured that the analysis would capture a SPOF of the RPS due to CCFs. Including all other interdivisional CCF combinations (e.g., A-B, ABC, CBD, etc.) would provide an increase in cut sets, but is beyond the scope of this case study. It is assumed acceptable to ignore the SPOFs due to CCFs within sub-divisions or modules.

6. RTS component and software diversity is ignored. All components that have identical function are assumed to be identical components (i.e., there is no diversity provided in the system). Limiting diversity simplifies the identification of CCFs and helps to direct the use of safety features, such as diversity, to mitigate CCFs. It should be noted that if diversity is known, it can be included. However, the analyst should refer to NUREG/CR-6303, NUREG/CR-7007, NUREG-KM-0009 and NUREG/CR-5485, as necessary, to make informed decisions regarding the adequacy of diversity and where to include CCFs in the FT [11] [12] [17] [18].

7. Each control action (e.g., trip signal) originates from an independent action within the controller rather than a single control action that is electrically or physically split to multiple destinations. This was done to follow specifically the STPA process, which assumes nothing of the physical connections between controllers. This provides the most unbiased controller requirements for designers to work with. Decisions regarding how to distribute or duplicate control actions can be a point of defense in a system design. STPA allows the designer to make decisions how best to defend the system after the fact. The trip signal from the bistable processors (BPs) to individual logic processors (LPs) is the one notable exception to this. These connections, not shown in the control structure, are assumed to be physically split and distributed to each LP. This simplified the analysis from 512 potential UCAs associated with the BPs to 128.

8. No bypassing function, maintenance work, or corrective actions are considered in the analysis.

9. The hazard analysis assumes the RTS is monitoring an NPP operating normally at 100% power, with the control rods completely withdrawn and full power to the turbines prior to the anticipated operational occurrence (AOO).

10. Individual sensor failures are grouped into one single basic event per division (e.g. division A sensor failure).

**STEP-1: Create a detailed representation of the SOI.**

The RTS is responsible for controlling the automatic insertion of reactor control rods into the core to bring the nuclear reaction to a shut-down state. For the sake of this hazard analysis, the RTS contains four main controllers capable of causing reactor shutdown: (1) the MCR; (2) the RSR; (3) the diverse protection system (DPS); and (4) the reactor protection system (RPS). The MCR and the RSR are manual components, while the DPS and RPS are automatic. The automatic control elements of the RTS maintain four divisional redundancy, while the manual components (assumed for this analysis) have no divisional redundancy. The case study is simplified by narrowing the analysis to module-level detail of the RPS and divisional or higher resolution for the remaining three parts of the RTS, as shown in Figure 6.

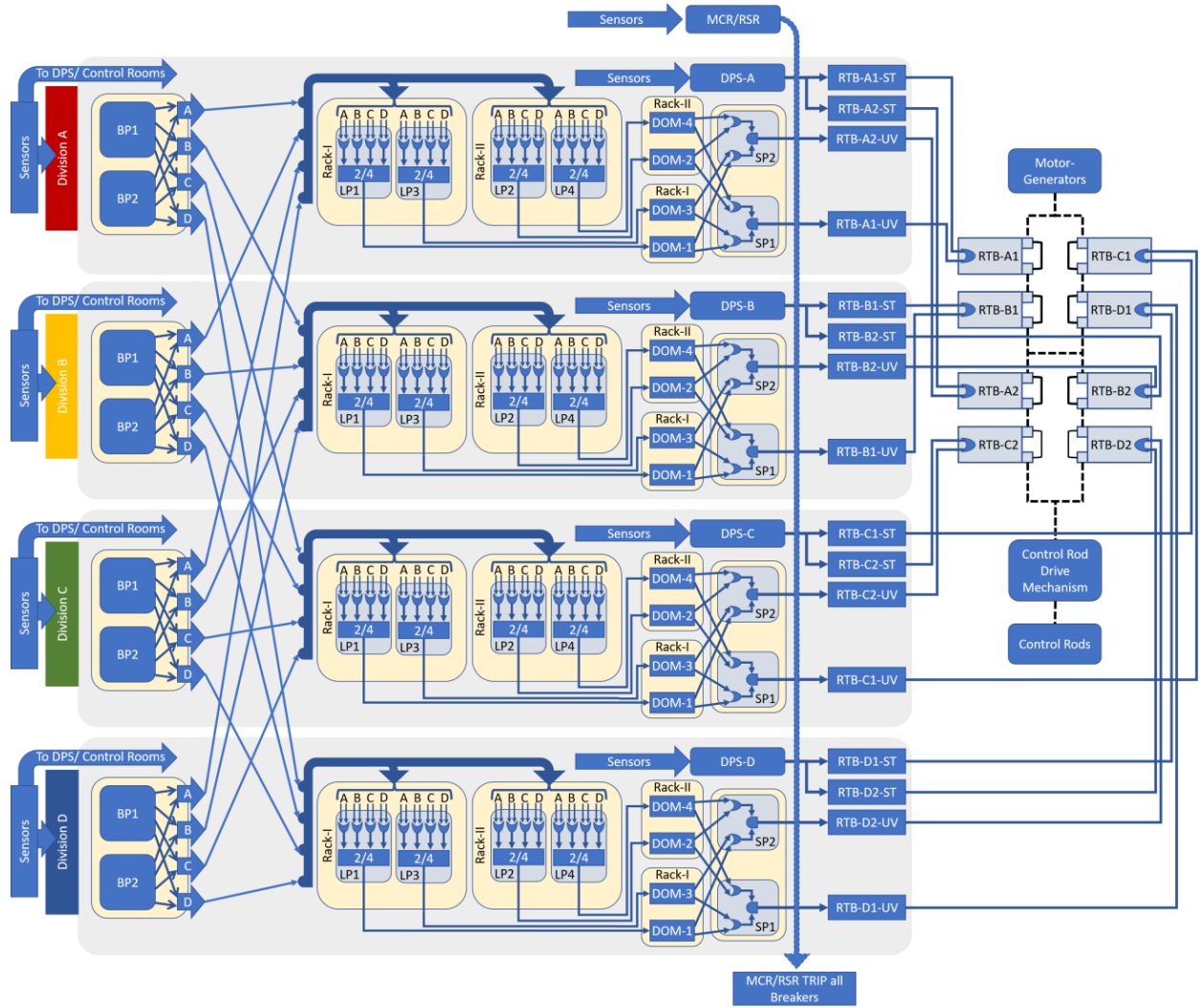

Figure 6. Detailed representation of the RTS.

**STEP-2: Develop a FT consisting of the hardware failures for a chosen function of the SOI.**

STEP-2A: The most significant function of the RTS is to trip or scram the reactor by rapidly inserting the control rods. The top event selected for the FT is therefore chosen to be that the RTS fails to trip the reactor during an AOO, which may be any event that can be anticipated for the reactor requiring shutdown via trip or scram.

STEP-2B: The FT is assembled based only on the hardware failures of the component of the RTS, according to the process described previously. The FT should also include hardware CCF basic events for every common or redundant hardware component of the RTS. Figure 7 shows a portion of the hardware-based FT. Note that the structure also includes the logic indicated by the system sketch.

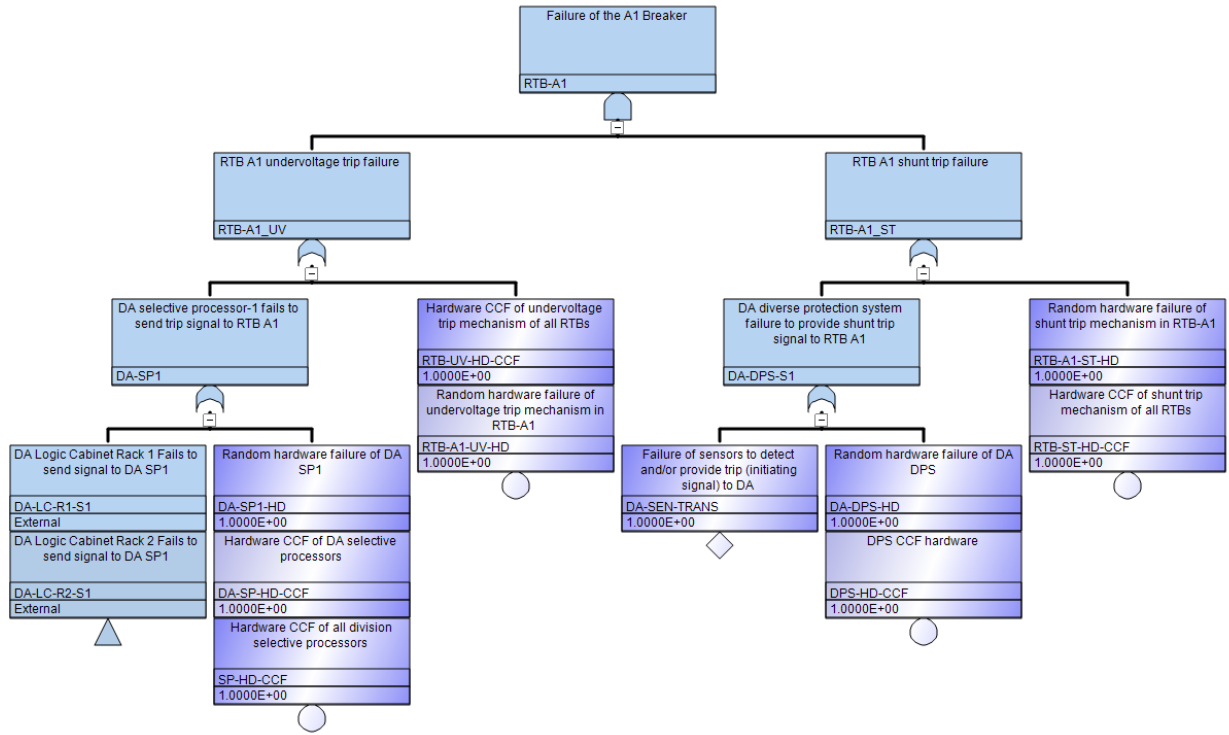

Figure 7. Portion of RTS FT showing hardware-type failures only. The U.S. NRC probabilistic risk assessment software SAPHIRE was used to construct the FT [34].

**STEP-3: Determine UCAs based on a redundancy-guided application of STPA.**

STEP-3A: In order to gain a clear understanding of the system, tables of losses and hazardous states that can lead to those losses are created. The RTS system is designed to prevent these hazardous states from occurring and ultimately prevent the losses in the table. The hazardous states demonstrate system level hazards and provide context for UCAs to be identified later. The following tables contain the losses and system-level hazards of the RTS.

Table 1: Major losses to be prevented [6]

| L1 | Human injury or loss of life |
|----|------------------------------|
| L2 | Environmental contamination  |
| L3 | Equipment damage             |
| L4 | Power generation             |
| L5 | Public perception            |

Table 2. Hazards which may lead to Losses [6]

| H1 | Reactor temperature too high (L1, L2, L3, L4, L5) |
|----|----------------------------------------------------|
| H2 | Equipment beyond limits (L1, L2, L3, L4, L5)       |
| H3 | Release of radioactive materials (L1, L2, L5)      |
| H4 | Reactor shutdown (L4, L5)                           |

STEP-3B: A redundancy-guided multi-layer control structure is created. Each layer of redundancy in the RTS is used to create a control structure diagram. The first layer of the RTS contains the redundancy for the trip function across four diverse controlling subsystems: MCR, RSR, DPS, and RPS. The second layer of redundancy is found in the multiple divisions of the DPS and the RPS. The next layers of the control structure come from the units and modules found in the RPS. The DPS sub-divisional redundancy is ignored to simplify the case study. Figure 8 shows the multi-layer control structure for division A. The remaining divisions for the case study are identical in format. Each structure indicates control actions and feedback in the system. The unknown control actions (CAx in Figure 8) are populated as they are discovered by layer.

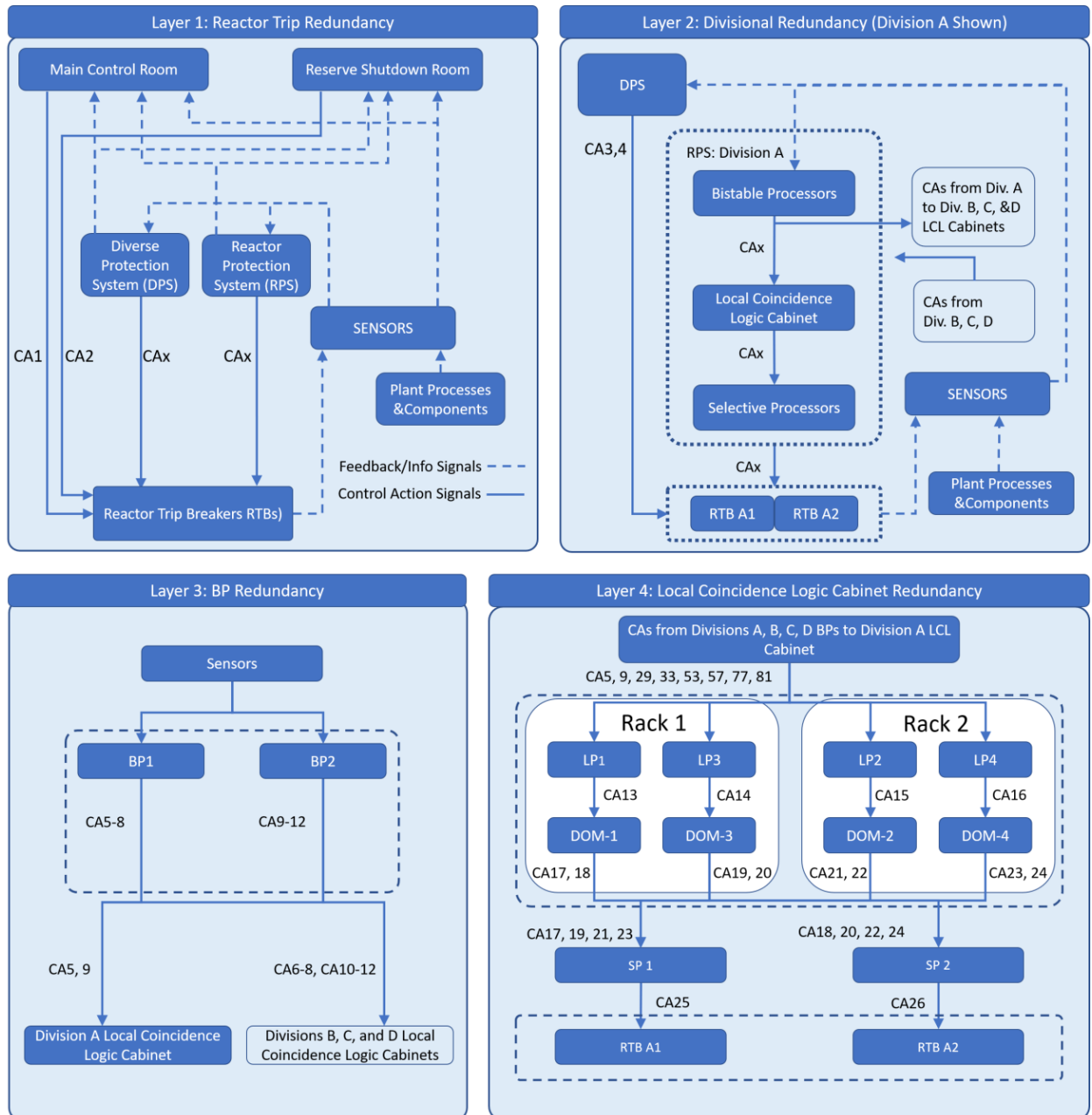

Figure 8. Redundancy-guided multi-layer control structure.

STEP-3C: The control actions found in each control structure are compiled in a table and analyzed according to the four categories of UCAs. Each UCA follows the format provided previously. Table 3 shows an example of the UCA table. It is assumed that Case D is not applicable as a trip command applied too long should not matter during an AOO. Also, it is assumed that a trip command cannot be stopped too soon as it is not a continuous controlling action.

Table 3. Examples of UCAs.

| Control Action (CA) | UCAa: CA is needed, but not given | UCAb: CA is Given, but not needed | UCAc: CA is given too early, too late, wrong order | UCAd: CA is applied too long or stopped too soon |
|---|---|---|---|---|
| CA18: DOM-1 demands SP1 to trip the reactor | UCA18a: DOM-1 does not provide trip command to SP1 during AOO [H1, H2, H3]. | UCA18b: DOM-1 provides trip command to SP1 when there is NO AOO [H4]. | UCA18c: DOM-1 provides trip command to SP1 after AOO has existed for some time [H1, H2, H3]. | UCA18d: Not applicable. |
| CA20: DOM-3 demands SP1 to trip the reactor | UCA20a: DOM-3 does not provide trip command to SP1 during AOO [H1, H2, H3]. | UCA20b: DOM-3 provides trip command to SP1 when there is NO AOO [H4]. | UCA20c: DOM-3 provides trip command to SP1 after AOO has existed for some time [H1, H2, H3]. | UCA20d: Not applicable. |
| Note: AOO: Anticipated Operational Occurrence; DOM: Digital Output Module; SP: Selective Processor. | | | | |

**STEP-4: Construct an integrated FT by adding applicable UCAs as basic events.**

The top event chosen for this case study is that the RTS fails to trip the reactor during an AOO. Based on this assumption, the appropriate UCAs are chosen (types: UCAa and UCAb) to populate the FT with software failures. Figure 9 shows the software failures added to the FT and update from Figure 7.

**STEP-5: Identify potential software CCFs based on duplicate or redundant UCAs within the FT.**

Within the FT, there are redundant or duplicate control actions that can be used to add additional software CCF basic events. The FT in Figure 9 shows the basic events of UCA type A and UCA type C, as these were deemed appropriate to add to the FT in the previous step. These UCAs have commonality across redundant divisions or units. For example, Division A Selective Processor 1 from Figure 9 is the same component as Division A Selective Processor 2 (unit redundancy) and the same component as Division B,C, and D Selective Processors (divisional redundancy). Each of these common components can be seen in Figure 6. Assuming all of these components are from same manufacturer, with identical software and functionality, they have the potential for a CCF. In the FT, under each category for software failures, the newly identified CCF events are added. CCFs in this model are assumed to occur within a single division and across all divisions. There are no subsets of CCFs between divisions (e.g., A-B, B-C, C-D etc.) Figure 9 shows the software-based CCFs that have been added to the FT.

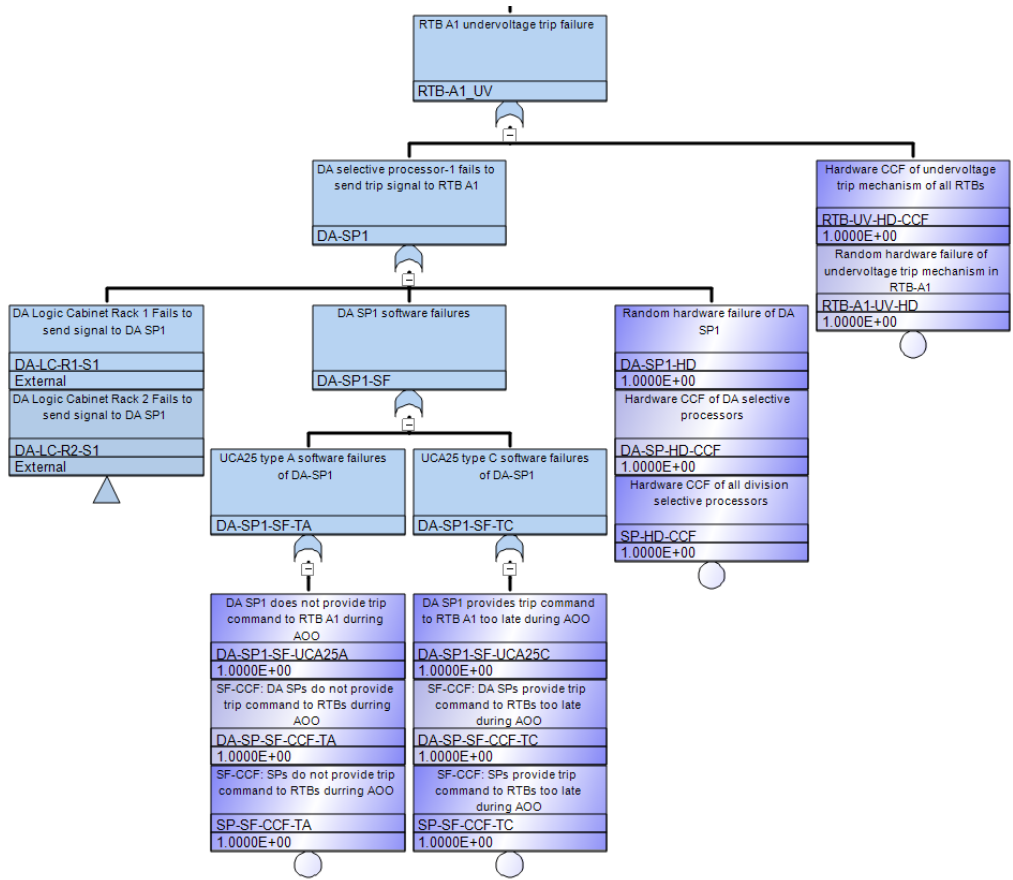

Figure 9. Portion of FT showing the UV trip failure of RTB A1 with software failures and CCFs added.

**STEP-6 Solve the FT for the minimal cut sets to determine potential SPOFs in the design.**

The minimal cut sets were found using the SAPHIRE software program (the same program used to create the FT). In order to evaluate and provide a list of cut sets without having failure data, the cut sets were listed and truncated based on order rather than by probability. For this case study, the number of basic events contributing to the failure of the top event is limited to a truncation of six or fewer basic events. Table 4 contains results for the FT for varying conditions: (1) complete RTS; (2) hardware only failures; (3) automatic trip functions only; and (4) RPS only.

Table 4. Cut set results.

| Truncation (order) | Full RTS model | RTS hardware only | Automatic trip only | RPS only |
|---|---|---|---|---|
| None | N/A | 15234 | N/A | N/A |
| 6 | 1,184,652 | - | 4,583,568 | N/A |
| 5 | 85788 | - | 1,038,956 | 328,355 |
| 4 | 468 | - | 13,1628 | 54,899 |
| 3 | 0 | - | 9,532 | 15,283 |
| 2 | 0 | - | 52 | 1,203 |
| 1 | 0 | - | 0 | 13 |

Based on the information shown in Table 4, only the final column has any SPOFs. This result was obtained by ignoring the diversity usually provided by the DPS, RSR, and MCR. Each of these could have also been evaluated singularly for SPOFs, but the case study purposefully simplified. Results for the RPS are given in Table 5. There are thirteen total SPOFs—five Hardware-based and eight software-based. All of them are due to CCFs (see footnote 5 in Section 3 regarding CCFs and SPOFs).

Table 5: First order cut sets or single points of failure for the RPS system, UV trip only.

| Number | Cut set | Description |
| --- | --- | --- |
| 1 | SP-HD-CCF | Selective processor hardware CCF. |
| 2 | LC-DOM-HD-CCF | Logic cabinet digital output module hardware CCF. |
| 3 | RTB-UV-HD-CCF | Reactor trip breaker undervoltage hardware CCF. |
| 4 | LC-BP-HD-CCF | Logic bistable processor hardware CCF. |
| 5 | LC-LP-HD-CCF | Logic cabinet logic processor hardware CCF |
| 6 | LC-LP-SF-CCF-TA | Logic cabinet logic processor software CCF type A. |
| 7 | LC-LP-SF-CCF-TC | Logic cabinet logic processor software CCF type C. |
| 8 | LC-DOM-SF-CCF-TA | Logic cabinet digital output module software CCF type A. |
| 9 | LC-DOM-SF-CCF-TC | Logic cabinet digital output module software CCF type C. |
| 10 | SP-SF-CCF-TA | Selective processor software CCF type A. |
| 11 | SP-SF-CCF-TC | Selective processor software CCF type C. |
| 12 | LC-BP-SF-CCF-TA | Logic cabinet bistable processor software CCF type A. |
| 13 | LC-BP-SF-CCF-TC | Logic cabinet bistable processor software CCF type C. |

**STEP-7: Identify and provide guidance to eliminate triggers of critical failures in the design including CCFs and SPOFs**.

The focus of this section is on the CCFs and SPOFs of the system. However, these techniques can be applied to any event of interest found in the cut set lists. The idea is to provide possible sources for their failure. The STPA Handbook indicates that the causes of UCAs can be grouped into two categories: (1) unsafe controller behaviors; and (2) inadequate feedback and/or other inputs [23]. Hardware-based failures are reasonably understood due largely because of industry experience [35]. Identifying causal factors for these failures is therefore based on historical failure-rate data. For software-based failures, the analysis can be more challenging and requires the researcher to consider the possible causes of unsafe controller behavior or inadequate feedback. The following is an example of these two categories applied to one of the software CCFs found in Table 5. Specifically, the chosen basic event is LC-BP-SF-CCF-TC (i.e., Logic Cabinet Bistable Processor CCF of Software Type C) corresponding to the failure of all the BPs to provide a trip command to each the logic cabinet of each division too late during an AOO.

- Causal factors due to category 1: unsafe controller behavior:

  Scenario: The nuclear reactor experiences an event meriting a reactor trip/scram. During this time, the BP should recognize the status of the plant and demand the reactor to trip. Processing delays within the BP result in the control action occurring too late. Software engineers should provide guidance as to the causes of processing delays, because shared software can lead to a potential CCF.

- Causal factor due to category 2: inadequate feedback

  Scenario: A BP may experience failure due to a lack of adequate feedback from the plant leading to the BP having an incorrect view of the status of the plant. For example, the BP relies on steam generator pressure information. This signal for steam generator pressure may be corrupt or incorrect, resulting in the BP failing to act appropriately. The BP may "think" the system is at an appropriate pressure and do nothing for some time before the pressure reaches a value corresponding to what the BP "thinks" is necessary to trip the reactor. Based on the assumption that all BPs have the same software, this would result in a CCF of the system. Of course, the CCF originates in faulty sensors, but also extends to the control action of the BPs and the subsequent UCA.

- Guidance for these two scenarios:

  For the first scenario regarding software processing delays, the use of diverse software processing as part of D3 may help at reduce the risk of having all controllers fail commonly. Thus, eliminating the risk of CCF even if the single failure mode still might occur. However, with a system that has multiple layers of redundancy and diversity, such as this RTS system, there will always be at least four diverse ways to shut down/trip/scram the reactor. Based on having four means to trip the reactor at the highest level make this system design, from a purely qualitative standpoint, relatively safe.

  For the second scenario, the proper use of D3 will help eliminate the common failure potential of shared software and hardware. In addition, the use of proper procedural testing and maintenance can ensure that the sensors will provide correct information to the BP. Diversity in sensor measurements can also help to ensure safety by providing a backup source of feedback to the BP. Both processes are already commonly employed by the NRC and industry. See NUREG/CR-6303, NUREG/CR-7007, NUREG-KM-0009 and NUREG/CR-5485 to make informed decisions regarding the adequacy of diversity and where to include CCFs in the FT [11] [12] [17] [18].

## 5. DISCUSSION

This paper has demonstrated an approach in the application of STPA and FTA for hazard analysis, as well as the identification of potential CCFs, SPOFs, and their causal factors. The treatment of redundancy by STPA was altered, making the capture of redundant features the central focus in both the creation of the control structure and the analysis of the UCAs. STPA was combined with FTA in a redundancy-guided way to provide insight to the relationships and failure modes of the digital RTS. RESHA identified 196 UCAs which resulted in 1,184,652 cut sets truncated to the sixth order. Future work will quantitatively refine the cut sets based on probability values. For now, these results can be qualitatively refined and grouped by specific subsections of the RTS. For the RPS, qualitative results indicated a total of thirteen potential CCFs. Determination of the triggers of these most critical hazards were discussed with a couple examples in the preceding section. It is recognized that a multidisciplinary team is needed in order to identify all relevant triggers.

## 6. CONCLUSIONS

This work has provided a means to overcome technical challenges faced by the nuclear industry for the implementation of digital I&C systems. In particular, RESHA provided a means to identify software-based interactions and potential CCFs in highly redundant, state-of-the-art digital I&C systems, by fully incorporating redundancy into the hazard analysis process. Embracing redundancy in the analysis allowed the work to meet its objectives in three ways: (1) this work defines a step-by-step approach for the hazard analysis of digital systems, that can help engineers efficiently make design and risk mitigation decisions by providing them a means to systematically identify the most critical CCFs and hazards of digital I&C systems; (2) RESHA identifies the critical hazards of a system, allowing utilities to effectively manage the cost of safety-rated digital I&C by strategically eliminating unnecessary design features; (3) RESHA provides a technical basis for reliability analysis by identifying crucial failure modes and qualitatively determining their effects on system vulnerability. Ultimately, RESHA helps improve the design of highly redundant digital I&C through a detailed qualitative hazard analysis.

## 7. ACKNOWLEDGEMENTS

article for publication, acknowledges that the U.S. Government retains a nonexclusive, paid-up, irrevocable, worldwide license to publish or reproduce the published form of this manuscript, or allow others to do so, for U.S. Government purposes. This work was also supported by consultations and contributions from Ken Thomas, and James Knudsen from INL, Andrew Clark and Adam Williams from Sandia National Laboratory, and Edward (Ted) Quinn from Technology Resources.